\documentclass[preprintnumbers,amsmath,amssymb]{revtex4}

\usepackage{graphicx}
\usepackage{dcolumn}
\usepackage{bm}

\begin{document}


\title{Image transfer through a chaotic channel by intensity correlations}
\author{Maria Bondani$^1$, Emiliano Puddu$^{1,2}$ and Alessandra Andreoni$^{1,2}$}
\email{Maria.Bondani@uninsubria.it}
\affiliation{$^1$Istituto Nazionale per la Fisica della Materia, INFM, Unit{\`a} di Como\\$^2$Dipartimento di Fisica e Matematica, Universit{\`a} degli Studi dell'Insubria\\Via Valleggio 11, 22100 Como, Italy}


\begin{abstract}
The three-wave mixing processes in a second-order nonlinear medium
can be used for imaging protocols, in which an object field is
injected into the nonlinear medium together with a reference field
and an image field is generated. When the reference field is
chaotic, the image field is also chaotic and does not carry any
information about the object. We show that a clear image of the
object be extracted from the chaotic image field by measuring the
spatial intensity correlations between this field and one Fourier
component of the reference. We experimentally verify this imaging
protocol in the case of frequency downconversion.
\end{abstract}

\maketitle

\section{Introduction}
Since the very beginning of its history, second-order ($\chi^{(2)}$)
nonlinear optical processes, also called three-wave mixing (TWM),
have been used to implement image processing. Initially the process
involved was frequency upconversion, namely second-harmonic
generation \cite{Midwinter68a, Midwinter68b, Firester69a,
Firester69b, Firester69c}. Though contemporarily proposed
\cite{Firester69b, Firester69c} image-processing in downconversion
took some more time to be realized \cite{Faris94, Devaux95a,
Devaux95b, Devaux95c, Gavrielides87, Avizonis77, Lefort96}.
The imaging capability of the nonlinear $\chi^{(2)}$ interactions
are connected with the spatial properties of the process, which in
turn reflect the presence of phase- and frequency-matching
conditions among the interacting light fields\cite{Kolobov}.
In the past years, we have investigated, both theoretically and
experimentally, the common features between image processing in TWM
and the Gabor holography \cite{Gabor49, JOSAB00, PRA02, JOSAB03},
that were first recognized by Firester \cite{Firester69a,
Firester69b, Firester69c}.
As in an holographic process, TWM imaging protocols require the
presence of a reference field, a role that can be played by any of
the interacting fields. The generated image thus depends on the
simultaneous presence of both the object and the reference fields:
this opens the possibility of performing conditional measurements
exploiting the correlations between the reference and the image
fields.
Imaging protocols based on correlations have been used in the
quantum domain, in which the generation of couples of entangled
photons by spontaneous downconversion and their detection by
coincidence techniques \cite{Serg_2}, allow to transfer spatial
information from one of the twin photons to the other \cite{Belin,
Souto_4, Serg_1, Barbosa}. In these protocols, often called ghost
imaging, the object is inserted on either signal or idler beyond the
crystal. Similar results have been obtained by using a classical
source of correlated single-photon pairs \cite{Benninck_1,
Benninck_2}.
In the many photon regime, it has been theoretically shown that the
same kind of image transfer can be implemented with both quantum
entangled \cite{Gatti_2} and classically correlated light
\cite{Cheng, Gatti_6, Magatti}.
One may alternatively place the object to be imaged on the beam
pumping the spontaneous parametric downconversion process
\cite{Abou}. Also in this case, images have been actually detected
by mapping suitable photon coincidences between the single-photon
pairs in the generated beams \cite{Pittman,Souto_5,Souto_6}.
\par
In this paper we exploit the correlation properties intrinsic to
$\chi^{(2)}$ processes to recover the image of an object transmitted
through a chaotic channel. In fact, if we use a chaotic reference
field for the TWM interaction, we obtain the generation of a chaotic
image, that does not contain any recognizable information on the
original object. Nevertheless, we will demonstrate that by
evaluating the intensity correlations between one of the Fourier
components of the reference field and the chaotic generated image,
we can recover a clean image of the object.
As shown in \cite{OLinpress}, this imaging protocol can be
considered a way to simulate many-photon quantum-imaging experiments
by using classical fields.
\par
The paper is organized as follows: in Section~\ref{s:holo} we
present the analytical description of the imaging process
implemented by a $\chi^{(2)}$ interaction in a nonlinear crystal and
and experimentally verify the theoretical predictions for a 2-D
object; in Section~\ref{s:chaotic} we show, both analytically and
experimentally, how to use the intensity correlations to recover the
image of the object in the case of a chaotic seed.
\section{Image transfer in parametric downconversion}\label{s:holo}
In this Section we present a full 3-D theory of three wave mixing
and apply it to the case of a seed field (at $\omega_1$) interacting
with an intense pump field (at $\omega_3>\omega_1$). The interaction
produces the amplification of seed field and the simultaneous
generation of a field at $\omega_2 =\omega_3-\omega_1$. This
interaction can be used for transferring an amplitude modulation
from the pump beam to the generated beam. We demonstrated that this
imaging process has holographic properties by showing that, when the
amplitude modulation is set by a 3D object, the wavefronts of the
generated field reconstruct a 3D image of the object\cite{OL00}. In
the experimental paragraph of the Section we verified the
theoretical expectations as to location and sizes of the images.
\subsection{Theory}\label{ss:theoholo}
\begin{figure}[hc]
\hspace{0mm}
\includegraphics[width=8cm]{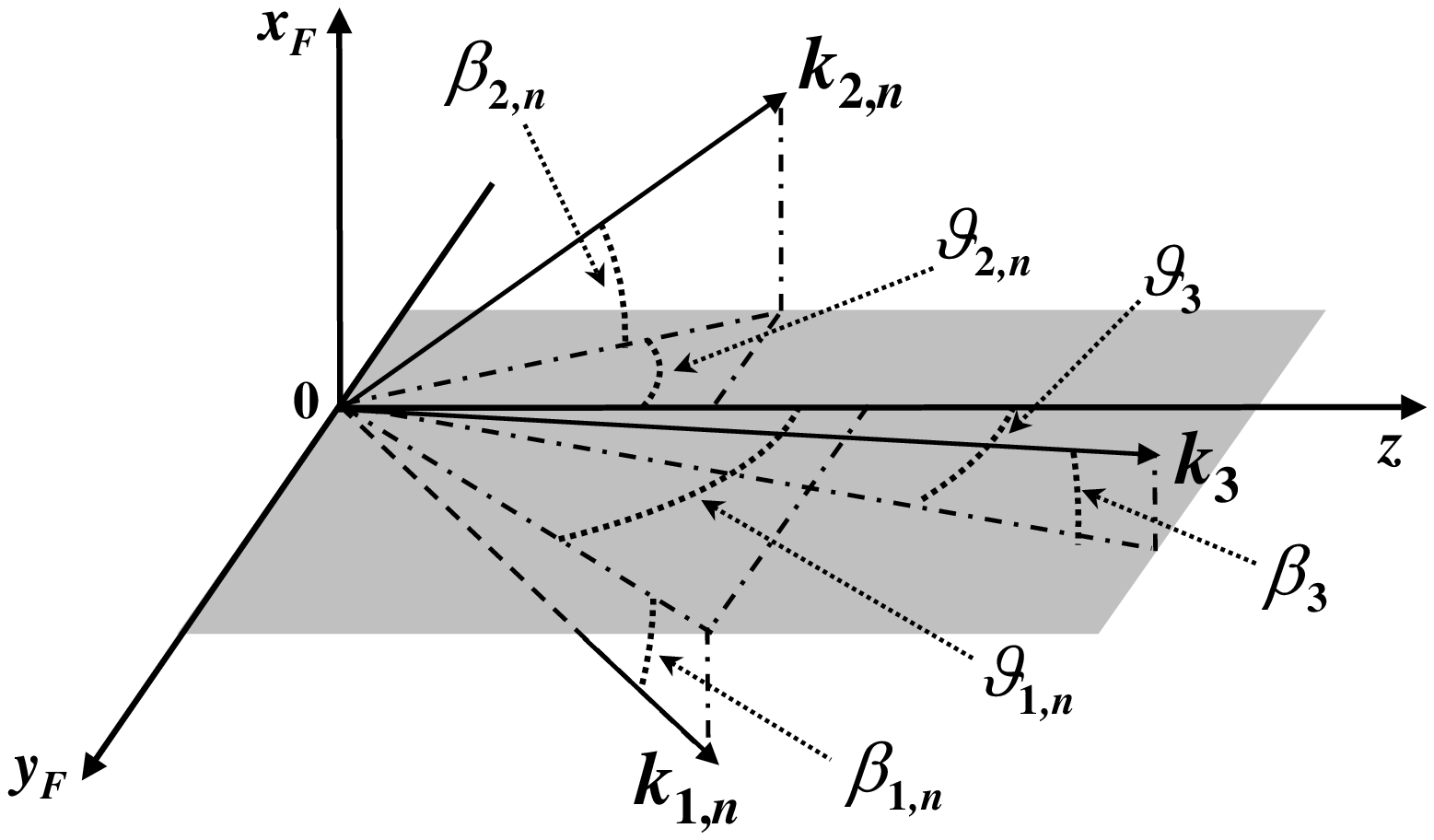}
\caption{Interaction inside the crystal. The shaded plane contains
the optical axis and the normal to the crystal entrance face
($z-$axis).} \label{f:interaz}
\end{figure}

According to the geometry of Fig. \ref{f:interaz}, we describe the
interaction inside a nonlinear crystal of the following three
amplitude-modulated plane-waves, propagating along ${\mathbf{k_j}}$
%

%
\begin{eqnarray}
{\mathbf E}_{1} \left({\mathbf r},t \right) &=& \frac{\hat{\mathbf
x}}{2}\sqrt {\frac{{2\eta _0 \hbar \omega _1 }}{{n_1 }}}\ a_{1}
\left( {\mathbf r} \right)\exp \left(- i {\mathbf{k_1}}\cdot
{\mathbf r}\right) + c.c.\nonumber\\
{\mathbf E}_{2} \left({\mathbf r},t \right) &=& \frac{\hat{\mathbf
x}}{2}\sqrt {\frac{{2\eta _0 \hbar \omega _2 }}{{n_2 }}}\ a_{2}
\left( {\mathbf r} \right)\exp \left(- i {\mathbf{k_2}}\cdot
{\mathbf r}\right) + c.c.\nonumber\\
{\mathbf E}_{3} \left({\mathbf r},t \right) &=& \frac{1}{2}\sqrt
{\frac{{2\eta _0 \hbar \omega_3}}{{n_3 }}}\ \left[ \hat{\mathbf y}\
a_{3y} \left( {\mathbf r} \right) + \hat {\mathbf z}\ a_{3z}
\left({\mathbf r} \right) \right]\ \exp \left(- i
{\mathbf{k_3}}\cdot {\mathbf r}\right) + c.c.\ ,\label{eq:fields}
\end{eqnarray}
where $n_j$ are the refraction indexes, $\eta_0$ is the vacuum
impedance and the optical frequencies, $\omega_j$, satisfy energy
matching ($\omega_3 =\omega_2+\omega_1$). In the "slowly varying
envelope" approximation, the system describing the second-order
interaction inside the crystal is
\begin{eqnarray}
{\mathbf{\hat{k}_1}}\cdot{\mathbf{\nabla}} a_{1} &=& i\left[ {g_ +
a_{3y} \left( {\mathbf r} \right) + g_ - a_{3z} \left( {\mathbf r}
\right)} \right]a_{2}^* \left( {\mathbf r} \right)\exp \left[
- i\Delta {\mathbf{k}}\cdot {\mathbf r}\right] \nonumber\\
{\mathbf{\hat{k}_2}}\cdot{\mathbf{\nabla}} a_{2} &=& i\left[ {g_ +
a_{3y} \left( {\mathbf r} \right) + g_ - a_{3z} \left( {\mathbf r}
\right)} \right]a_{1}^* \left( {\mathbf r} \right)\exp \left[
- i\Delta {\mathbf{k}}\cdot {\mathbf r}\right]\nonumber\\
{\mathbf{\hat{k}_3}}\cdot{\mathbf{\nabla}} a_{3y} &=& ig_ + a_{1}
\left( {\mathbf r} \right)a_{2} \left( {\mathbf r}
\right)\exp \left[i\Delta {\mathbf{k}}\cdot {\mathbf r}\right] \nonumber\\
{\mathbf{\hat{k}_3}}\cdot{\mathbf{\nabla}} a_{3y} &=& ig_ - a_{1}
\left({\mathbf r} \right)a_{2} \left( {\mathbf r} \right)\exp
\left[i\Delta {\mathbf{k}}\cdot {\mathbf r}\right] \ ,
\label{eq:systemvect}
\end{eqnarray}
where $g_{\pm}$ are coupling coefficients \cite{Nikogosyan97} and we
have defined the phase mismatch vector as $ \Delta {\mathbf{k}} =
\mathbf{k_3} -\mathbf{k_{2}}-\mathbf{k_{1}}$.
\par\noindent
In order to analytically solve (\ref{eq:systemvect}), one of the
interacting fields must be taken as non evolving during the
interaction: we are interested in the case of non-evolving
transverse field $a_3 \left(\mathbf r\right) = a_3 \left(0 \right)$.
system (\ref{eq:systemvect}) becomes
\begin{eqnarray}
{\mathbf{\hat{k}_1}}\cdot{\mathbf{\nabla}} a_{1} &=& ig\ a_{3}
\left( {0} \right)a_{2}^* \left( {\mathbf r} \right)\exp \left[
- i\Delta {\mathbf{k}}\cdot {\mathbf r}\right] \nonumber\\
{\mathbf{\hat{k}_2}}\cdot{\mathbf{\nabla}} a_{2} &=& ig\ a_{3}
\left( {0} \right)a_{1}^* \left( {\mathbf r} \right)\exp \left[ -
i\Delta {\mathbf{k}}\cdot {\mathbf r}\right]\ ,
\label{eq:system2vecteff}
\end{eqnarray}
where $g$ is a new effective interaction coefficient.
The solution of (\ref{eq:system2vecteff}) is given by
\begin{eqnarray}
a_{1} \left({\mathbf r} \right) &=& \left\{a_{1} \left(0
\right)\left[\cosh\left(Q\frac{\widehat{\Delta \mathbf{k}}}{2}\cdot
\mathbf{r}\right) + i\frac{\Delta k}{Q} \sinh\left(Q
\frac{\widehat{\Delta \mathbf{k}}}{2}\cdot \mathbf{r}\right)
\right]+\right.\nonumber\\
&&+\left. a_{2}^* \left(0 \right) \frac{2i g a_3 \left( {0}
\right)}{\left(\widehat{\Delta \mathbf{k}}\cdot \mathbf{\hat{k}_1}
\right)Q} \sinh\left(Q\frac{\widehat{\Delta \mathbf{k}}}{2}\cdot
\mathbf{r}\right)\right\}\exp\left(-i\frac{\widehat{\Delta
\mathbf{k}}}{2}\cdot \mathbf{r}\right)
\nonumber\\
a_{2} \left({\mathbf r} \right) &=& \left\{a_{1}^* \left(0
\right)\frac{2i g a_3 \left( {0} \right)}{\left(\widehat{\Delta
\mathbf{k}}\cdot \mathbf{\hat{k}_2} \right)Q}
\sinh\left(Q\frac{\widehat{\Delta \mathbf{k}}}{2}\cdot
\mathbf{r}\right)
+\right.\nonumber\\
&&+\left. a_{2}\left(0
\right)\left[\cosh\left(Q\frac{\widehat{\Delta \mathbf{k}}}{2}\cdot
\mathbf{r}\right) + i\frac{\Delta k}{Q}
\sinh\left(Q\frac{\widehat{\Delta \mathbf{k}}}{2}\cdot
\mathbf{r}\right) \right] \right\}\exp\left(-i\frac{\widehat{\Delta
\mathbf{k}}}{2}\cdot \mathbf{r}\right)\ ,\label{eq:solDk}
\end{eqnarray}
where
\begin{eqnarray}
Q = \left(\frac{4 g^2 \left|a_3 \left( {0}
\right)\right|^2}{\left(\widehat{\Delta \mathbf{k}}\cdot
\mathbf{\hat{k}_1} \right)\left(\widehat{\Delta \mathbf{k}}\cdot
\mathbf{\hat{k}_2} \right)} - \Delta k ^2\right)^{1/2}\ .
\label{eq:defQ}
\end{eqnarray}
When the wavevectors satisfy the phase-matching condition ($\Delta
\mathbf{k} = 0$), solutions (\ref{eq:solDk}) become:
\begin{eqnarray}
a_{1}\left({\mathbf r} \right) &=& a_{1} \left(0
\right)\cosh\left(\frac{g \left|a_3 \left( {0}
\right)\right|}{\mathbf{\hat{b}}\cdot \mathbf{\hat{k}_{1,2}}}
\mathbf{\hat{b}}\cdot \mathbf{r}\right)+ a_{2}^* \left(0 \right)
\frac{a_3 \left( {0} \right)}{\left|a_3 \left( {0} \right)\right|}
\sinh\left(\frac{g \left|a_3 \left( {0}
\right)\right|}{\mathbf{\hat{b}}\cdot \mathbf{\hat{k}_{1,2}}}
\mathbf{\hat{b}}\cdot \mathbf{r}\right)
\nonumber\\
a_{2}\left({\mathbf r} \right) &=& ia_{1}^* \left(0 \right)\frac{a_3
\left( {0} \right)}{\left|a_3 \left( {0} \right)\right|}
\sinh\left(\frac{g \left|a_3 \left( {0}
\right)\right|}{\mathbf{\hat{b}}\cdot \mathbf{\hat{k}_{1,2}}}
\mathbf{\hat{b}}\cdot \mathbf{r}\right)+ a_{2}\left(0
\right)\cosh\left(\frac{g \left|a_3 \left( {0}
\right)\right|}{\mathbf{\hat{b}}\cdot \mathbf{\hat{k}_{1,2}}}
\mathbf{\hat{b}}\cdot \mathbf{r}\right)\ \label{eq:solPM}
\end{eqnarray}
where $\mathbf{b}$ is along the bisector of  $\psi$, the angle
$\mathbf{\hat{k}_1}$-to-$\mathbf{\hat{k}_2}$, {\textit{i.e.}}
$\mathbf{b} = (\mathbf{\hat{k}_1}+\mathbf{\hat{k}_2})/2$ . Since
$\cos\psi$ can be written as
\begin{eqnarray}
\cos\psi = \sin \beta_{1}\sin \beta_{2} + \cos \beta_{1}\cos
\beta_{2}\cos\left(\vartheta_{1}-\vartheta_{2}\right)\ ,
\label{eq:psi}
\end{eqnarray}
we find that $b =\cos(\psi/2)(=\mathbf{\hat{b}}\cdot
\mathbf{\hat{k}_{1,2}})$ to be used in (\ref{eq:solPM}). Thus
$(\mathbf{\hat{b}}\cdot \mathbf{r})/(\mathbf{\hat{b}}\cdot
\mathbf{\hat{k}_{1,2}}) = f(\vartheta,\beta) r$, where
\begin{eqnarray}
f(\vartheta,\beta)=\frac{\sin \beta_{1}+\sin \beta_{2}+
\cos\beta_{1}\sin \vartheta_{1}+\cos \beta_{2}\sin \vartheta_{2} +
\cos \beta_{1}\cos \vartheta_{1}+\cos \beta_{2}\cos
\vartheta_{2}}{2\cos^2\psi/2}\  \label{eq:broverbk12}
\end{eqnarray}
is a pure geometrical factor. In the following we will consider the
particular case of $a_{2} \left(0\right) = 0$ and $a_1(0)\neq 0$ in
the weak-conversion approximation ($g \left|a_3 \left( {0}
\right)\right|f(\vartheta,\beta)\ r\ll 1$), for which the solution
(\ref{eq:solPM}) can be written as
\begin{eqnarray}
a_{1}\left({\mathbf r} \right) &\simeq& a_{1} \left(0\right)
\nonumber\\
a_{2}\left({\mathbf r} \right) &=& i g\ f(\vartheta,\beta)\ r\
a_{1}^*\left( {0} \right) a_3\left(0 \right)\ ,\label{eq:solappr}
\end{eqnarray}
that is the amplification of field $a_1$ is almost negligible and
the generated field $a_2$ is linearly dependent on both the pump
field  $a_3$ and the incident field  $a_1$.
\par\noindent
Note that (\ref{eq:solappr}) gives the values of the evolved fields
at any position ${\mathbf r}$ inside the crystal, given a starting
point ${\mathbf r} = 0$ at the entrance face. Different field
components starting at different initial points evolve independently
and solution (\ref{eq:solappr}) can be thus generalized for any
transverse field distribution incident on the crystal. In
particular, if field $a_1(0)$ is a plane wave and the non-evolving
pump field $a_3(0)$ is amplitude modulated, (\ref{eq:solappr}) shows
that the modulation is transferred to the generated field $a_2$.
\subsection{Image formation}\label{ss:imageholo}
We will now explicitly calculate the "image" field generated on
$a_2$ given an "object" field $a_3$ in a particular propagation
scheme that will be useful for experimental purposes.

\begin{figure}[hc]
\hspace{0mm}
\includegraphics[width=8cm]{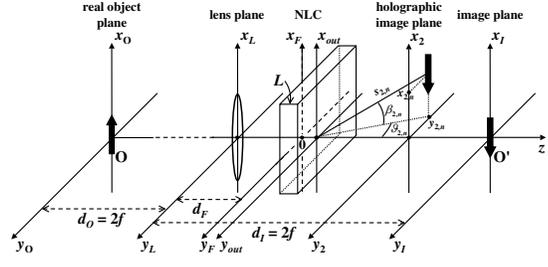}
\caption{Propagation scheme; NLC, nonlinear crystal.} \label{f:prop}
\end{figure}

According to Fig.~\ref{f:prop}, we consider a modulation of the pump
field $a_3^{O}(x_O, y_O)$ on the object plane $(x_O, y_O)$. We
insert a converging lens (focal length $f$) on the plane $(x_L,
y_L)$ located at a distance $d$ from the object plane and at a
distance $d_F$ from the crystal entrance face, plane $(x_F, y_F)$.
The field distribution at the entrance face of the crystal in the
presence of an infinite ideal lens is given by \cite{Goodman88}
\begin{eqnarray}
a_3\left({\mathbf r_F}\right) &=& - \frac{k_3^2}{4\pi^2 d_O
d_F}\int\int dx_O dy_O a_3^{O}\left(x_O, y_O\right)\exp
\left[-i\frac{k_3\left(x_F^2 + y_F^2\right)}{2d_F}\right]
\exp\left[-i\frac{k_3\left(x_O^2 +
y_O^2\right)}{2d_O}\right]\nonumber\\
&\times&\int\int dx_L dy_L \exp
\left[-i\frac{k_3}{2}\left(\frac{1}{d_O}
+\frac{1}{d_F}-\frac{1}{f}\right) \left(x_L^2 + y_L^2\right)\right]
\exp \left\{-ik_3\left[\left(\frac{x_O}{d_O} +\frac{x_F}{d_F}\right)
x_L + \left(\frac{y_O}{d_O} +\frac{y_F}{d_F}\right) y_L\right]
\right\}\ . \label{propag}
\end{eqnarray}
By evaluating the second integral and letting $d_O = 2f$ ($2f-2f$
system) we simplify the expression as
\begin{eqnarray}\label{eq:propag3}
a_3 \left({\mathbf r_F}\right) &=& \frac{k_3}{2\pi i d}
\exp\left[{i\frac{k_3}{2d}\left(x_F^2 +
y_F^2\right)}\right]\nonumber\\
&\times& \int dx_O dy_O a_3^{O}\left(x_O, y_O\right)
\exp\left[{i\frac{k_3}{2d}\frac{f-d}{f}\left(x_O^2 +
y_O^2\right)}\right] \exp\left[{i\frac{k_3}{d}\left(x_F x_O + y_F
y_O\right)}\right]\ ,
\end{eqnarray}
where $d\equiv 2f-d_F$ is the distance between the crystal and the
image plane of the lens.
We now let the field impinging on each point of the crystal entrance
face interact according to (\ref{eq:solappr}) by setting
$a_3\left(0\right)\rightarrow a_3\left({\mathbf r_F}\right)$.
According to (\ref{eq:fields}), the spatial amplitude of the
generated field ${\mathbf E}_{2}$ becomes
\begin{eqnarray}
E_{2}\left({\mathbf r_{out}}\right) &\propto& a_{2} \left( {\mathbf
r_{out}} - {\mathbf r_{F}}\right)\exp \left[- i
{\mathbf{k_2}}\cdot\left(
{\mathbf r_{out}}- {\mathbf r_{F}}\right)\right]\nonumber\\
&\propto& i g r f( \vartheta,\beta)\ a_{1}^*(0)\  a_{3}\left(
{\mathbf r_F}\right)\exp \left[- i {\mathbf{k_2}}\cdot\left(
{\mathbf r_{out}}- {\mathbf r_{F}}\right)\right]\ ,
\label{eq:fieldE2}
\end{eqnarray}
being $z_{out}-z_F = L$ the crystal depth and $r=|{\mathbf
r_{out}}-{\mathbf r_{F}}|$. In the limit of non-evolving pump, we
can take substitute in (\ref{eq:fieldE2}) $a_3\left({\mathbf
r_{F}}\right)$ with $a_3\left({\mathbf r_{out}}\right)$.
Now we let field ${\mathbf E}_{2}$ freely propagate from plane
$\left(x_{out}, y_{out}, z_{out} \right)$ to plane $\left(x_{2},
y_{2}, z_{2}\right)$ along its direction. By neglecting refraction
at the crystal interfaces, the field will travel a distance
$s_{2}=(z_2-z_{out})/(\cos\vartheta_{2} \cos\beta_{2})$ (see
Fig.~\ref{f:prop}). By inserting (\ref{eq:propag3}) into
(\ref{eq:fieldE2}) and applying free propagation, we get\\
\begin{eqnarray}
E_{2}\left({\mathbf r_{2}}\right) &\propto& i g r f(
\vartheta,\beta) \frac{k_2}{2\pi i s_{2}}\frac{k_3}{2\pi i d}\
a_{1}^*(0)  \exp \left[-i\frac{k_2\left(x_{2}^2 +
y_{2}^2\right)}{2s_{2}}\right]
\nonumber\\
&\times&\exp \left[-ik_2 \left(s_{2}+\cos \beta_{2}\cos
\vartheta_{2} L\right)\right]\exp \left[ ik_2 \left( \sin
\beta_{2} x_F + \cos \beta_{2} \sin \vartheta_{2} y_F \right)\right]\nonumber\\
&\times&\int \left\{\int {a_3^{O}\left(x_O, y_O\right)
\exp\left[{i\frac{k_3}{2d}\frac{f-d}{f}\left(x_O^2 +
y_O^2\right)}\right] \exp\left[{i\frac{k_3}{d}\left(x_{out} x_O +
y_{out} y_O\right)}\right] dx_O dy_O}\right\}\nonumber\\
&\times&\exp \left[ -ik_2\left( \sin \beta_{2} x_{out}
+ \cos \beta_{2} \sin \vartheta_{2} y_{out}\right) \right]\nonumber\\
&\times&\exp \left[i\frac{k_2}{s_{2}}\left(x_2 x_{out} + y_2
y_{out}\right) \right] \exp \left[i\frac{k_3\left(x_{out}^2 +
y_{out}^2\right)}{2d}\right]dx_{out} dy_{out}\ .
\label{eq:fieldE2prop2}
\end{eqnarray}
By exchanging the integrations and neglecting the quadratic term
inside the last integral we get\\
\begin{eqnarray}
E_{2}\left({\mathbf r_{2}}\right) &\propto& i g r f(
\vartheta,\beta) \frac{k_2}{2\pi i s_{2}}\frac{k_3}{2\pi i d}\
a_{1}^*(0)  \exp \left[-i\frac{k_2\left(x_{2}^2 +
y_{2}^2\right)}{2s_{2}}\right]\nonumber\\
&\times&\exp \left[-ik_2 \left(s_{2}+\cos \beta_{2}\cos
\vartheta_{2} L\right)\right]\exp \left[ ik_2 \left( \sin
\beta_{2} x_F + \cos \beta_{2} \sin \vartheta_{2} y_F \right)\right]\nonumber\\
&\times&\int\left\{\int \exp \left[-i\left(k_2 \sin \beta_{2}
-\frac{k_2}{s_{2}} x_2  -\frac{k_3}{d} x_O\right) x_{out} -i \left(
k_2 \cos \beta_{2} \sin \vartheta_{2} -\frac{k_2}{s_{2}}
y_2-\frac{k_3}{d}y_O\right)y_{out}\right] dx_{out}
dy_{out}\right\}\nonumber\\
&\times&a_3^{O}\left(x_O, y_O\right)
\exp\left[i\frac{k_3}{2d}\frac{f-d}{f}\left(x_O^2 +
y_O^2\right)\right] dx_O dy_O\nonumber\\
&=& i g r f(\vartheta,\beta) \frac{k_2}{2\pi i s_{2}}\frac{k_3}{2\pi
i d}\ a_{1}^*(0)  \exp \left[-i\frac{k_2\left(x_{2}^2 +
y_{2}^2\right)}{2s_{2}}\right]
\nonumber\\
&\times&\exp \left[-ik_2 \left(s_{2}+\cos \beta_{2}\cos
\vartheta_{2} L\right)\right]\exp \left[ ik_2 \left( \sin
\beta_{2} x_F + \cos \beta_{2} \sin \vartheta_{2} y_F \right)\right]\nonumber\\
&\times&4\pi^2\int\delta\left(k_2 \sin \beta_{2} -\frac{k_2}{s_{2}}
x_2 -\frac{k_3}{d}x_O, k_2 \cos \beta_{2} \sin \vartheta_{2}
-\frac{k_2}{s_{2}} y_2-\frac{k_3}{d}y_O\right)\nonumber\\
&\times&{a_3^{O}\left(x_O, y_O\right)
\exp\left[{i\frac{k_3}{2d}\frac{f-d}{f}\left(x_O^2 +
y_O^2\right)}\right] dx_O dy_O}\nonumber\\
&=& i g r f( \vartheta,\beta) \frac{k_2}{2\pi i s_{2}}\ a_{1}^*(0)
\exp \left[-i\frac{k_2\left(x_{2}^2 + y_{2}^2\right)}{2s_{2}}\right]
\frac{k_3}{2\pi i d}\nonumber\\
&\times&\exp \left[-ik_2 \left(s_{2}+\cos \beta_{2}\cos
\vartheta_{2} L\right)\right]\exp \left[ ik_2 \left( \sin
\beta_{2} x_F + \cos \beta_{2} \sin \vartheta_{2} y_F \right)\right]\nonumber\\
&\times&4\pi^2\left(\frac{d}{k_3}\right)^2\int
\delta\left(\frac{d}{k_3}k_2 \sin \beta_{2}
-\frac{d}{k_3}\frac{k_2}{s_{2}} x_2 - x_O, \frac{d}{k_3} k_2 \cos
\beta_{2} \sin \vartheta_{2}
-\frac{d}{k_3}\frac{k_2}{s_{2}} y_2-y_O\right)\nonumber\\
&\times&{a_3^{O}\left(x_O, y_O\right)
\exp\left[{i\frac{k_3}{2d}\frac{f-d}{f}\left(x_O^2 +
y_O^2\right)}\right] dx_O dy_O}\nonumber\\
&=& -i g r\ 4\pi^2\left(\frac{d}{k_3}\right)^2 \frac{k_2k_3 }{4\pi^2
d} \frac{f( \vartheta,\beta)}{s_{2}}\ a_{1}^*(0) \exp \left[-i\frac{k_2\left(x_{2}^2 + y_{2}^2\right)}{2s_{2}}\right]\nonumber\\
&\times&\exp \left[-ik_2 \left(s_{2}+\cos \beta_{2}\cos
\vartheta_{2} L\right)\right]\exp \left[ ik_2 \left( \sin
\beta_{2} x_F + \cos \beta_{2} \sin \vartheta_{2} y_F\right)\right]\nonumber\\
&\times&\exp\left[{i\frac{k_2^2}{k_3 s_{2}^2} \frac{d(f-d)}{2
f}\left[\left(s_{2}\sin \beta_{2} - x_2\right)^2 + \left(s_{2}\cos
\beta_{2} \sin \vartheta_{2} - y_2\right)^2\right]}\right]\nonumber\\
&\times& a_3^{O}\left(\frac{k_2}{k_3}\frac{d}{s_{2}}\left(s_{2}\sin
\beta_{2} - x_2\right),\frac{k_2}{k_3}\frac{d}{s_{2}}\left(s_{2}\cos
\beta_{2} \sin \vartheta_{2} - y_2\right)
\right)\nonumber\\
&=& -i g r\frac{k_2}{k_3} \frac{d}{s_{2}}f( \vartheta,\beta)
\exp\left[i\Phi\left({\mathbf r_{2}}\right)\right] \ a_{1}^*(0)\nonumber\\
&\times& a_3^{O}\left(\frac{k_2}{k_3}\frac{d}{s_{2}}\left(s_{2}\sin
\beta_{2} - x_2\right),\frac{k_2}{k_3}\frac{d}{s_{2}}\left(s_{2}\cos
\beta_{2} \sin \vartheta_{2} - y_2\right) \right)\ ,
\label{eq:fieldE2prop3}
\end{eqnarray}
where we have introduced the phase factor $\Phi\left({\mathbf
r_{2}}\right)$.
By writing all the coefficients in a single complex function $C$ we
can summarize our result as
\begin{eqnarray}
{E}_{2}\left({\mathbf r_{2}}\right) = C a_{1}^*(0)a_3^O
\left(\frac{k_2}{k_3}\frac{d}{s_{2}}\left(\bar{x}_{2}-
x_2\right),\frac{k_2}{k_3}\frac{d}{s_{2}}\left(\bar{y}_{2} -
y_2\right)\right)\ , \label{eq:resul}
\end{eqnarray}
where we have defined
\begin{eqnarray}
\bar{x}_{2} = s_{2}\sin \beta_{2}\ ;\ \bar{y}_{2} = s_{2}\cos
\beta_{2} \sin \vartheta_{2}\ . \label{eq:defxy}
\end{eqnarray}
If we locate the plane $\left(x_{2}, y_{2}, z_{2}\right)$ so that
${s_{2} \simeq k_2/{k_3}{d}}$ \cite{JOSAB04}, we can simplify the
expression as
\begin{eqnarray}
{\mathbf E}_{2}\left({\mathbf r_{2}}\right) = C a_{1}^*(0)a_3^{O}
\left(\bar{x}_{2}- x_2,\bar{y}_{2} - y_2\right)\ . \label{eq:result}
\end{eqnarray}
The result in (\ref{eq:result}) shows that with this choice of the
propagation configuration, the field distribution on a plane
satisfying the holographic relation among the distances is equal to
the field distribution on the object plane, provided a double
inversion of the coordinates and a shift in position which depends
on the propagation direction ${\mathbf E}_1$. Note that the
insertion of the imaging lens on the plane $(x_L,y_L)$ was necessary
to obtain a real image \cite{JOSAB04}.
%

\subsection{Experiment}\label{ss:expholo}
\begin{figure}[hc]
\hspace{0mm}
\includegraphics[width=8cm]{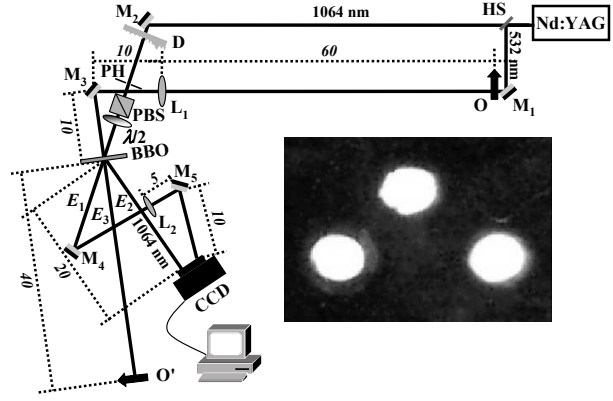}
\caption{Experimental setup: HS, harmonic separator; D, diffusing
plate (removable); M$_{1-5}$, mirrors. Lens L$_1$ images O into O$'$
through a $2f-2f$ system. Lens L$_2$ realizes the Fourier transform
of field $E_1$ on the CCD sensor. Inset: image taken with an optical
microscope of the copper sheet containing the three holes used to
produce the object-field modulation. The diameter of the holes was
$\sim 200\ \mu$m.} \label{f:setup}
\end{figure}

\begin{figure}[hc]
\hspace{0mm}
\includegraphics[width=8 cm]{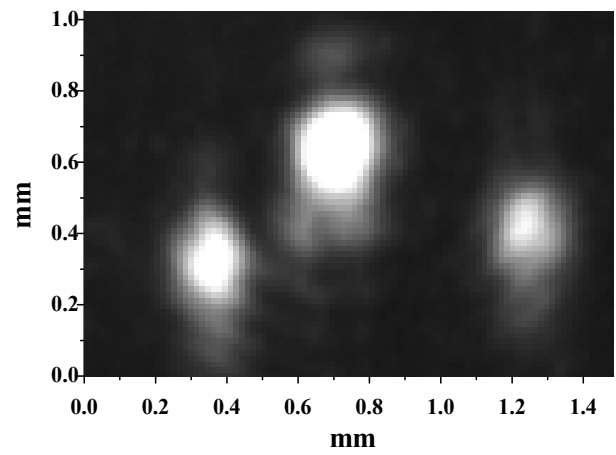}
\caption{Single-shot holographic image detected in the plane
$(x_2,y_2,z_2)$.} \label{f:holo}
\end{figure}

We experimentally verified the imaging properties of the interaction
by using the setup depicted in Fig.~\ref{f:setup}
The nonlinear crystal was a type I $\beta$-BaB$_2$O$_4$ crystal
(BBO, cut angle $32^\mathrm{o}$, $10$ mm $\times~10$ mm $\times~4$
mm, Fujian Castech Crystals).
The object field was produced by locating on the plane $(x_O,y_O)$ a
copper sheet with three holes ($\sim 256$ $\mu$m diameter, see inset
of Fig.~\ref{f:setup}) as the mask producing object O. The imaging
lens located in the plane $(x_L,y_L)$ at a distance $d_O=$ 60 cm
from the object plane had a focal length $f=300 $ mm so as to
realize a $2f-2f$ system.
The fields ${\mathbf E}_1$ and ${\mathbf E}_3$ entering the crystal
were obtained from the fundamental ($\lambda_1$ = 1064 nm) and the
second harmonics ($\lambda_3$ = 532 nm) outputs of a Nd:YAG laser
(10 Hz repetition rate, $7$-ns pulse duration, Spectra-Physics)thus
$\lambda_2$ = 1064 nm. The BBO crystal was located at a distance
$d_F = 20$ cm from the lens, so that the image plane of the lens
resulted to be at a distance $d = 40$ cm beyond the crystal. A
system made of a polarizing beam splitter plus a half-wavelength
plate was used to obtain a ordinarily polarized field ${\mathbf
E}_1$. For the present experiment, the diffuser D in
Fig.~\ref{f:setup} was removed from the setup. The sensor of a CCD
camera (Dalsa CA-D1-256T, 16 $\mu$m $\times$ 16 $\mu$m pixel area,
12 bits resolution, operated in progressive scan mode) was located
in the detection plane $(x_2,y_2,z_2)$. The distance of the CCD from
BBO was chosen to be $s_2 = 20$ cm, so as to satisfy $s_2 = d\
k_2/k_3$. In Fig.~\ref{f:holo} we show the image of the holes as
detected by the CCD camera. Note that in agreement with
(\ref{eq:result}) the transverse dimensions of the image were equal
to those of the object.
%

%
\section{Chaotic image-transfer in parametric downconversion}\label{s:chaotic}
In this Section we consider the same interaction seeded by a chaotic
field  ${\mathbf E}_1$, which amounts to realize the same image
transfer process of above through a chaotic channel. We will
demonstrate that in this "chaotic" situation the imaging system
gives no straightforward information on the object. Nevertheless,
the intensity fluctuations intrinsic to the chaotic light can be
profitably used to implement a protocol for correlated imaging that
allows the recovery of the object image.
\subsection{Theory}\label{ss:theochaotic}
We now suppose that the seed field ${\mathbf E}_{1}$ in
(\ref{eq:fields}) can be written as an incoherent superposition $N$
plane waves having random complex amplitudes, $a_{1,n}$, and wave
vectors, ${\mathbf k}_{1,n}$, with random directions but equal
amplitudes, $k_{1,n}=2\pi/\lambda_1$
\begin{eqnarray}
{E}_{1} \left({\mathbf r},t \right) \propto \frac{\hat{\mathbf
x}}{2}\sqrt {\frac{{2\eta _0 \hbar \omega _1 }}{{n_1 }}}\
\sum_{n=1}^N a_{1,n} \left( {\mathbf r} \right) \exp \left(- i
{\mathbf{k_{1,n}}}\cdot {\mathbf r}\right)\ .\label{eq:E1chao}
\end{eqnarray}
According to solution (\ref{eq:solappr}), field ${\mathbf E}_{1}$
does not evolve inside the crystal so that its spatial part can be
written as ${E}_{1}\left({\mathbf r}_{out}\right)\simeq
{E}_{1}\left({\mathbf r}_{F}\right)$.
Due to the chaotic nature of field ${\mathbf E}_{1}$, we expect that
also its Fourier transform is chaotically distributed. In fact, if
we insert a lens of focal length $\tilde{f}$ on the path of field
${\mathbf E}_{1}$ at a distance $\tilde{d}$ from the exit plane of
the nonlinear crystal, so as to have the Fourier plane on
$\left(x_{1}, y_{1}, z_{1}\right)$ we get\\
\begin{eqnarray}
{E}_{1}\left({\mathbf r}_{1}\right)&\propto&\frac{k_1}{2\pi i f}
\exp \left[-i\frac{k_1\left(x_1^2 +
y_1^2\right)}{2\tilde{f}}\left(1-\frac{\tilde{d}}{\tilde{f}}\right)\right]\nonumber\\
&\times&\int {{E}_{1}\left({\mathbf r}_{out}\right) \exp
\left[i\frac{k_1}{\tilde{f}}\left(x_1 x_{out} + y_1 y_{out}\right)
\right] dx_{out} dy_{out}}\ . \label{eq:fourierE1}
\end{eqnarray}
By substituting the definition of ${\mathbf E}_{1}$
\begin{eqnarray}
{E}_{1}\left({\mathbf r}_{1}\right) &\propto& \sum_{n=1}^N a_{1,n}
\frac{k_1}{2\pi i \bar{f}} \exp\left[-i\frac{k_1\left(x_1^2 +
y_1^2\right)}{2\tilde{f}}
\left(1-\frac{\tilde{d}}{\tilde{f}}\right)\right]\nonumber\\
&\times& \int \exp \left(- i {\mathbf{k_{1,n}}}\cdot {\mathbf
r}_{out}\right) \exp\left[i\frac{k_1}{\tilde{f}}\left(x_1 x_{out} +
y_1 y_{out}\right)
\right] dx_{out} dy_{out}\nonumber\\
&=& \sum_{n=1}^N C'_n a_{1,n} \exp \left[-i\frac{k_1\left(x_1^2 +
y_1^2\right)}{2\tilde{f}}\left(1-\frac{\tilde{d}}{\tilde{f}}\right)\right]\nonumber\\
&\times&\delta\left( \frac{k_1}{\tilde{f}}(x_1-\tilde{f}\sin
\beta_{1,n}),\frac{k_1}{\tilde{f}}(y_1-\tilde{f}\cos \beta_{1,n}\sin
\vartheta_{1,n})\right)\ ,\label{eq:fourier}
\end{eqnarray}
which means that if the amplitudes $a_{1,n}$ have random values, the
intensity distribution on the plane has the form of a speckle field:
\begin{eqnarray}
{I}_{1}\left({\mathbf r}_{1}\right) \propto \left|{
E}_{1}\left({\mathbf r}_{1}\right)\right|^2 =\sum_{n=1}^N
\left|C''_n\right|^2\left|{a}_{1,n}\right|^2
\delta\left(x_1-\tilde{f}\sin \beta_{1,n},y_1-\tilde{f}\cos
\beta_{1,n}\cos \vartheta_{1,n}\right)\ , \label{eq:int1}
\end{eqnarray}
Inside the nonlinear medium, each of the spatial Fourier components
of the seed field that is phase matched with the pump field
generates an independent contribution to ${\mathbf E}_{2}$ according
to (\ref{eq:solappr}). The overall field at the crystal output face
is given by
\begin{eqnarray}
{E}_2\left({\mathbf r}_{out} \right)\propto \sum_{n=1}^{N} i g r_n
f(\vartheta_n,\beta_n) a_{1,n}^*(0) a_3({\mathbf r}_{F})
\exp\left[-i\ {\mathbf k}_{2,n}\cdot\left({\mathbf r}_{out}-{\mathbf
r}_{F}\right)\right]\ ,\label{eq:E2chao}
\end{eqnarray}
where each wavevectors ${\mathbf k}_{2,n}$ are assumed to satisfies
the phase-matching conditions ${\mathbf k}_{3}={\mathbf
k}_{1,n}+{\mathbf k}_{2,n}$. As obvious, this assumption cannot be
satisfied for all wavevectors ${\mathbf k}_{1,n}$, thus setting an
angular limitation to the effectiveness of the interaction. The
result (\ref{eq:resul}) can thus be used to evaluate each of the
terms in (\ref{eq:E2chao}) to find the $N$ contributions to field
${\mathbf E}_2$
\begin{eqnarray}
{E}_{2}\left({\mathbf r}_{2,n}\right) =
C_{n}a_{1,n}^*a_3^{O}\left(\frac{k_2}{k_3}\frac{d}{s_{2,n}}\left(\bar{x}_{2,n}-
x_2\right),\frac{k_2}{k_3}\frac{d}{s_{2,n}}\left(\bar{y}_{2,n} -
y_2\right)\right)\ , \label{eq:resulCHAO}
\end{eqnarray}
where we have defined
\begin{eqnarray}
\bar{x}_{2,n} = s_{2,n}\sin \beta_{2,n}\ ;\ \bar{y}_{2,n} =
s_{2,n}\cos \beta_{2,n} \sin \vartheta_{2,n}\ . \label{eq:defxyCHAO}
\end{eqnarray}
If the effective angular spread of the wavevectors ${\mathbf
k}_{2,n}$ is not too broad, we can assume that ${s_{2,n}\ \simeq{d}
k_2/{k_3}}$  for all components and that the corresponding $N$
images form on the same plane. Thus from (\ref{eq:defxyCHAO}) we can
write
\begin{eqnarray}
{E}_{2}\left({\mathbf r}_{2}\right) = \sum_{n=1}^N C_{n}
a_{1,n}^*a_3^{O}\left(\bar{x}_{2,n}- x_2,\bar{y}_{2,n} - y_2\right)\
. \label{eq:resultCHAO}
\end{eqnarray}
Note that also the the field on plane $(x_2,y_2,z_2)$ results to be
the sum of $N$ holographic images of the pump field ${\mathbf E}_3$,
reconstructed in $N$ different transverse locations.
If we now take into account the random nature of the amplitudes
$a_{1,n}$, we obtain that the sum in (\ref{eq:resultCHAO}) is as
incoherent as that in (\ref{eq:E1chao}). We can thus write the
detected intensity of field $E_2$ in the plane $(x_2,y_2,z_2)$ as
\begin{eqnarray}
{I}_{2}\left({\mathbf r}\right) \propto \left|{E}_{2}\left({\mathbf
r}_{2}\right)\right|^2 = \sum_{n=1}^N
\left|C_{n}\right|^2\left|a_{1,n}\right|^2
\left|a_3^{O}\left(\bar{x}_{2,n}- x_2,\bar{y}_{2,n} -
y_2\right)\right|^2\ , \label{eq:int2}
\end{eqnarray}
This result means that no information about the object field is any
more detectable in a direct way.
Nevertheless, we can use the correlation properties between fields
${\mathbf E_1}$ and ${\mathbf E_2}$ introduced by the nonlinear
interaction in the crystal to recover the image of the object. To do
this, we select a single position $(x_1,y_1)$ in the plane of the
Fourier transform of ${\mathbf E_1}$, that, according to
(\ref{eq:int1}), corresponds to choosing a single plane wave having
amplitude $a_{1,j}$. We then calculate the correlations of the
temporal intensity fluctuations between field ${\mathbf E_1}$ and
field ${\mathbf E_2}$ we get\\
\begin{eqnarray}
\langle\Delta{I}_{1}\Delta{I}_{2}\rangle &=&
\langle{I}_{1}{I}_{2}\rangle-\langle{I}_{1}\rangle\langle{I}_{2}\rangle
\nonumber\\
&=& \langle\left|{a}_{1,j}\right|^2\sum_{n=1}^N
\left|C_{n}\right|^2\left|a_{1,n}\right|^2
\left|a_3^{O}\left(\bar{x}_{2,n}-
x_2,\bar{y}_{2,n} - y_2 \right)\right|^2\rangle \nonumber\\
&-& \langle\left|{a}_{1,j}\right|^2\rangle\langle\sum_{n=1}^N
\left|C_{n}\right|^2\left|a_{1,n}\right|^2
\left|a_3^{O}\left(\bar{x}_{2,n}-
x_2,\bar{y}_{2,n} - y_2 \right)\right|^2\rangle\nonumber\\
&=& \sum_{n=1}^N\langle\left|{a}_{1,j}\right|^2
\left|C_{n}\right|^2\left|a_{1,n}\right|^2
\left|a_3^{O}\left(\bar{x}_{2,n}-
x_2,\bar{y}_{2,n} - y_2 \right)\right|^2\rangle \nonumber\\
&-& \sum_{n=1}^N\langle\left|{a}_{1,j}\right|^2\rangle\langle
\left|C_{n}\right|^2\left|a_{1,n}\right|^2
\left|a_3^{O}\left(\bar{x}_{2,n}-
x_2,\bar{y}_{2,n} - y_2 \right)\right|^2\rangle \nonumber\\
&=&\sum_{n=1}^N
\left|C_{n}\right|^2\left|a_3^{O}\left(\bar{x}_{2,n}-
x_2,\bar{y}_{2,n} - y_2 \right)\right|^2
\left(\langle\left|{a}_{1,j}\right|^2 \left|a_{1,n}\right|^2\rangle
- \langle\left|{a}_{1,j}\right|^2\rangle\langle
\left|a_{1,n}\right|^2\rangle \right)\nonumber\\
&=& \sum_{n=1}^N
\left|C_{n}\right|^2\left|a_3^{O}\left(\bar{x}_{2,n}-
x_2,\bar{y}_{2,n} - y_2 \right)\right|^2
\sigma^2\left(\left|{a}_{1,n}\right|^2\right)
\delta_{j,n}\nonumber\\
&=&\left|C_{j}\right|^2\sigma^2\left(\left|{a}_{1,j}\right|^2\right)\left|a_3^{O}\left(x_{2,j}-
x_2,y_{2,j} - y_2 \right)\right|^2, \label{eq:corr1}
\end{eqnarray}
where $\sigma^2(x) = <x^2>-<x>^2$ is the variance. In deriving
(\ref{eq:corr1}) we used the fact that the coefficients $C_{n}$ do
not depend on the amplitudes ${a}_{1,n}$  (see
(\ref{eq:fieldE2prop3}) and (\ref{eq:resul})) and that the pump
field amplitude $a_3$ at each sample of the statistical ensemble.
The result in (\ref{eq:corr1}) shows that the correlation function
between the image on field $E_2$ and a single point on the Fourier
plane of field $E_1$ can reconstruct the image of the object encoded
on field $E_3$.
\subsection{Experiment}\label{ss:expchaotic}
For the experimental verification of the results of the previous
section, we used the same setup in Fig.~\ref{f:setup}, modified by
introducing the light diffuser $D$ on the seed beam $E_1$. The
diffuser, a ground-glass wheel, was moved from shot to shot of the
laser in order to obtain the temporal statistics needed to evaluate
the correlation function (\ref{eq:corr1}). A portion of the chaotic
field ${\mathbf E}_1$ emerging from the diffuser was selected with
an iris (PH, in Fig.~\ref{f:setup}) of $\sim$ 8 mm diameter and then
filtered in polarization with a polarizing beam splitter, PBS, and a
half-wave plate. Lens L$_2$ ($\tilde{f} = 15$ cm) provides the
Fourier transform of ${\mathbf E}_1$ on the plane $(x_1,y_1,z_1)$.
The detection planes of $(x_1,y_1,z_1)$ and $(x_2,y_2,z_2)$ were
made to coincide on the sensor of the same CCD camera so that each
signal occupies half sensor.
\par
\begin{figure}[hc]
\hspace{0mm}
\includegraphics[width=8 cm]{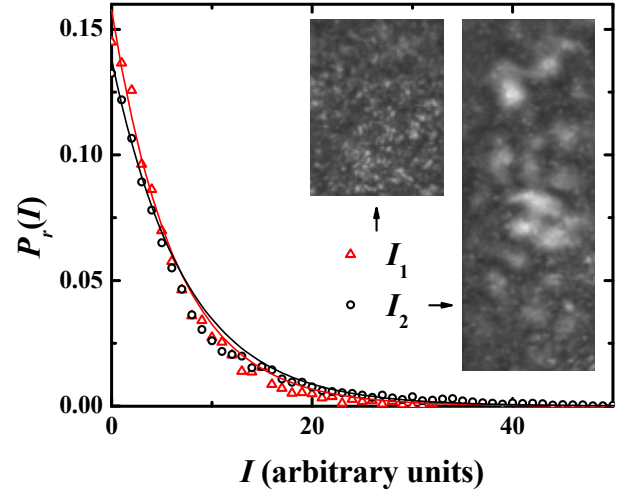}
\caption{Probability distributions, $P_{\bf r}({I}_{1,2})$  of the
intensity recorded by the different CCD pixels for a single shot,
relative to the Fourier transform of the seed field $E_1$ and to the
generated field $E_2$. Insets: single-shot intensity maps of $I_1$
and $I_2$.} \label{f:space}
\end{figure}

\begin{figure}[hc]
\hspace{0mm}
\includegraphics[width=8 cm]{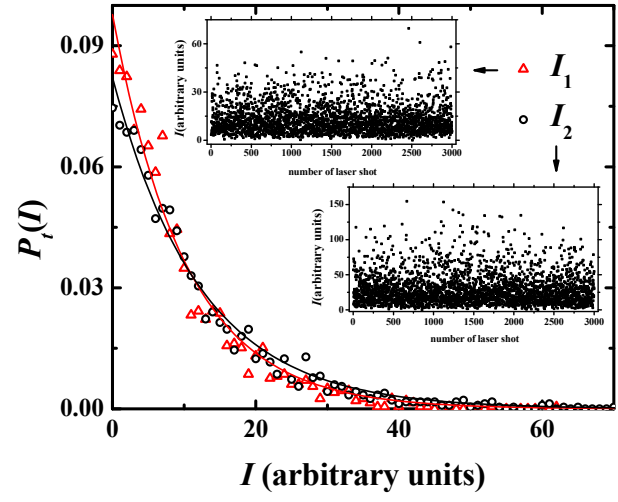}
\caption{Probability distributions, $P_{t}({I}_{1,2})$ of the
intensity of a selected CCD pixel recorded for many successive laser
shots. Insets: temporal traces of the intensities.} \label{f:time}
\end{figure}

\noindent
First of all, we checked the chaotic nature of both the seed field
$E_1$ and the generated field $E_2$ both in space and in time. A
characteristic of chaotic light is to have an intensity obeying
thermal distribution, $P(I)=\exp(I/<I>)/<I>$. We thus measured the
spatial probability distributions, $P_{\bf r}({I}_{1})$ and $P_{\bf
r}({I}_{2})$, of the intensity recorded by the different CCD pixels
for a single shot, relative to the Fourier transform of the seed
field $E_1$ and to the generated field $E_2$.
Figure~\ref{f:space} shows the results for the spatial distributions
along with the intensity maps used to evaluate them (insets). From
these results we can see that both the spatial distributions can be
fitted by a thermal distribution and that the intensity map relative
to the generated field $E_2$ has no memory of the field modulation
(the same three holes as before) imposed on the pump.
We also
measured the temporal (over many repetitions of the laser pulse)
probability distributions, $P_{t}({I}_{1})$ and $P_{t}({I}_{2})$, of
the intensity recorded by choosing a single pixel of the CCD in the
map of $E_1$ and $E_2$ and recording the intensity values at each
laser shot. Also in this case the intensity probability
distributions are well fitted by thermal distributions (see
Fig.~\ref{f:time}). The temporal traces of the intensities of the
selected pixels are shown in the insets of the figure.
\par
\begin{figure}[hc]
\hspace{0mm}
\includegraphics[width=8 cm]{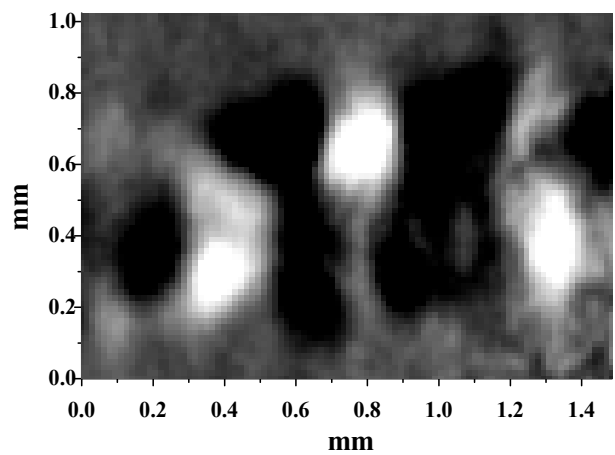} \caption{Map of
$G\left({I}_{1,j},{I}_{2}(x_2,y_2)\right)$ evaluated on 1000 shots}
\label{f:correl}
\end{figure}

\noindent
Once established the correspondence of our experimental setup with
the requirements of the theory, we evaluated the correlation
function (\ref{eq:corr1}) over 1000 shots by taking the whole map
$I_2\left(x_2,y_2\right)$ and by selecting the value of a single
pixel in the intensity map of $I_1$.
In Fig. \ref{f:correl} we show the resulting reconstructed image
(map of $G\left({I}_{1,j},{I}_{2}(x_2,y_2)\right)$): the similarity
in the quality of this image compared with that obtained without
diffuser (see Fig.~\ref{f:holo}) is really impressive, in particular
if the reconstructed image is compared with any of the single-shot
intensity maps $I_2\left(x_2,y_2\right)$, see for instance the inset
of Fig.~\ref{f:space}.
\section{Conclusions}\label{s:concl}
In conclusion, we have demonstrated that the spatial intensity
correlation properties of the downconversion process can be used to
recover a selected image from a chaotic ensemble of holographic
images. Note that theoretical results similar to those in Section
\ref{ss:theoholo} and \ref{ss:theochaotic} would be found for any
choice of the object and reference fields among the three
interacting ones. The image recovered by
$G\left({I}_{1,j},{I}_{2}(x_2,y_2)\right)$ fulfils the holographic
properties of the difference-frequency generated hologram that would
be obtained by using the single plane-wave {\bf E}$_{1,j}$ as the
seed/reference field. We expect that the method also works in the
case of an unseeded process, in which no reference field enters the
crystal. In this case any twin beam in the parametric fluorescence
cone would play the role of our reference and image fields and our
intensity correlation protocol should provide an \textit{a
posteriori} selection of a single holographic image. Note that the
plane on which this recovered image forms would depend on the choice
of the plane-wave component of the twin party used as the reference.
\section{Acknowledgements}
The authors thanks A. Gatti (I.N.F.M., Como) for stimulating
discussions, I.N.F.M. (PRA CLON) and the Italian Ministry for
University Research (FIRB n. RBAU014CLC$\_$002) for financial
support.

\newpage
\newpage
\newpage
\newpage
\newpage
\newpage
\newpage
\newpage

\end{document}